\documentclass[footinbib,aps,pra,reprint,superscriptaddress]{revtex4-2}

\usepackage{amsmath,amssymb,braket,upgreek,bm,verbatim}
\usepackage{graphicx,tabularx,xcolor} 
\definecolor{darkblue}{rgb}{0,0,0.5}
\usepackage{hyperref}
\hypersetup{
colorlinks=true,
linkcolor=black,
filecolor=blue,
citecolor=darkblue,  
urlcolor=black,
}

\usepackage[absolute]{textpos}
\usepackage{xcolor}
\usepackage{multirow}

\usepackage{amsmath,amssymb,bbm,bm}
\usepackage{mathtools}
\usepackage{braket}
\usepackage{amsmath}
\usepackage{relsize}
\usepackage{comment}
\usepackage{csquotes}

\usepackage{accents}

\makeatletter
\ams@newcommand{\iiiiint}{\DOTSI\protect\MultiIntegral{5}}
\renewcommand{\MultiIntegral}[1]{%
  \edef\ints@c{\noexpand\intop
    \ifnum#1=\z@\noexpand\intdots@\else\noexpand\intkern@\fi
    \ifnum#1>\tw@\noexpand\intop\noexpand\intkern@\fi
    \ifnum#1>\thr@@\noexpand\intop\noexpand\intkern@\fi
    \ifnum#1>4 \noexpand\intop\noexpand\intkern@\fi 
    \noexpand\intop
    \noexpand\ilimits@
  }%
  \futurelet\@let@token\ints@a
}
\makeatother

\usepackage{stackengine}

\newcommand{\appropto}{\mathrel{\vcenter{
  \offinterlineskip\halign{\hfil$##$\cr
    \propto\cr\noalign{\kern2pt}\sim\cr\noalign{\kern-2pt}}}}}

\begin{document}
\title{Ultrafast dynamic beam steering with optical frequency comb arrays}

\author{Suparna Seshadri}
\thanks{These authors contributed equally to this work. \\Contact email: sseshad@purdue.edu, wangjie181122@gmail.com}
\author{Jie Wang}
\thanks{These authors contributed equally to this work. \\Contact email: sseshad@purdue.edu, wangjie181122@gmail.com}
\affiliation{School of Electrical and Computer Engineering, Purdue University, West Lafayette, Indiana 47907, USA}
\author{Andrew M. Weiner}
\affiliation{School of Electrical and Computer Engineering, Purdue University, West Lafayette, Indiana 47907, USA}

\begin{abstract}

Efficient spatiotemporal control of optical beams is of paramount importance in diverse technological domains. Conventional systems focusing on quasi-static beam control demand precise phase or wavelength tuning for steering. This work presents a time-efficient solution for dynamic beam steering, emphasizing high-duty-cycle operation with fast scan rates, and eliminating the need for 
{active tuning} of the beam direction. We achieve 100\%-duty-cycle scans at a rate of $\sim$9.8 GHz within an angular range of $\sim$1$^\circ$. Furthermore, leveraging the dispersion characteristics of a virtually imaged phased array (VIPA), we devise a broadband source array that seamlessly transitions from continuous-angular steering to pulsed discrete-angular operation, unlocking possibilities for high-sensitivity angle-, range-, and time-resolved imaging. We also elucidate the adaptability of integrated photonic designs incorporating wavelength-selective switches and spectral dispersers, for enabling a versatile on-chip realization of the proposed beam steering schemes.




\end{abstract}
\maketitle
\vspace{-3mm}
\section*{Introduction}
\vspace{-3mm}

The significance of optical beam steering is underscored across various cutting-edge technologies, spanning applications including light detection and ranging~{(LIDAR)}\cite{lin2022high,kim2021nanophotonics}, free-space communication~\cite{poulton2019long, polkoo2019imaging}, and virtual/augmented reality displays~\cite{chen2016beam}. 
Conventional macromechanical scanners are bulky, slow, and lack mechanical durability, while more compact microelectromechanical systems (MEMS) still suffer from vulnerability to the environment with the moving parts, and liquid crystal-based systems are limited by the long response times. Overcoming these limitations is crucial for achieving improved autonomous navigation, enhanced data rates in free-space communication, and increased portability with faster frame rates for display technologies. In this direction, recent years have witnessed a rapid evolution in nanophotonics-based beam steering devices. Chip-scale designs, including optical phased arrays (OPAs) \cite{guo2021integrated,heck2017highly,berini2022optical}, spatial light modulators (SLMs) \cite{ li2019phase, panuski2022full}, and active phase gradient metasurfaces \cite{berini2022optical}, have successfully achieved quasi-static beam control by precisely manipulating the light phase from an array of wavelength-scale elements, each independently controlled. Despite these developments, challenges emerge when scaling the elements in the array, leading to the need for intricate control systems for phase tuning and increasing the fabrication complexity.

There have also been efforts exploring wavelength-dependent steering utilizing grating antennas, often integrated with one-dimensional OPA structures to realize two-dimensional steering~\cite{van2009off, xiao2005optical, im2020silicon}. Additionally, advancements in beam control dependent on acoustic wavelength have been demonstrated using Brillouin scattering, opening avenues for on-chip frequency-angular resolving LiDAR \cite{li2023frequency}.

While various investigations employ external phase control or wavelength tuning for beam direction scanning, there has been limited progress in the development of schemes for dynamic spatiotemporal beam steering. In our work, we explore dynamic steering schemes with the potential to achieve rapid scan rates without requiring active control of phase or wavelength. Unlike a phase-gradient array requiring active phase tuning to steer the beam, a recent demonstration of a frequency-gradient metasurface~\cite{shaltout2019spatiotemporal} introduced time-dependent phase differences between the emitters. This facilitated a continuous and automatic sweep of the beam direction at a periodic rate determined by the frequency increment between the emitters in the array. Despite these advancements, demonstrations so far have achieved a notably low duty cycle for the spatiotemporal scan--that is, the active scanning duration is under 0.5\% of the total scan period~\cite{shaltout2019spatiotemporal, li2023picosecond}. This arises due to the spectral disperser's limited resolution relative to the line-spacing of the optical frequency comb employed. This limitation is also evident in other demonstrations aiming to synthesize quasi-static spatiotemporal beams, where phase and amplitude control is restricted to spatially overlapping groups of comb lines following spectral dispersion~\cite{chen2022synthesizing,yessenov2022space,kondakci2017diffraction}.

\begin{figure*}
  \centering
  \includegraphics[width=0.94\textwidth]{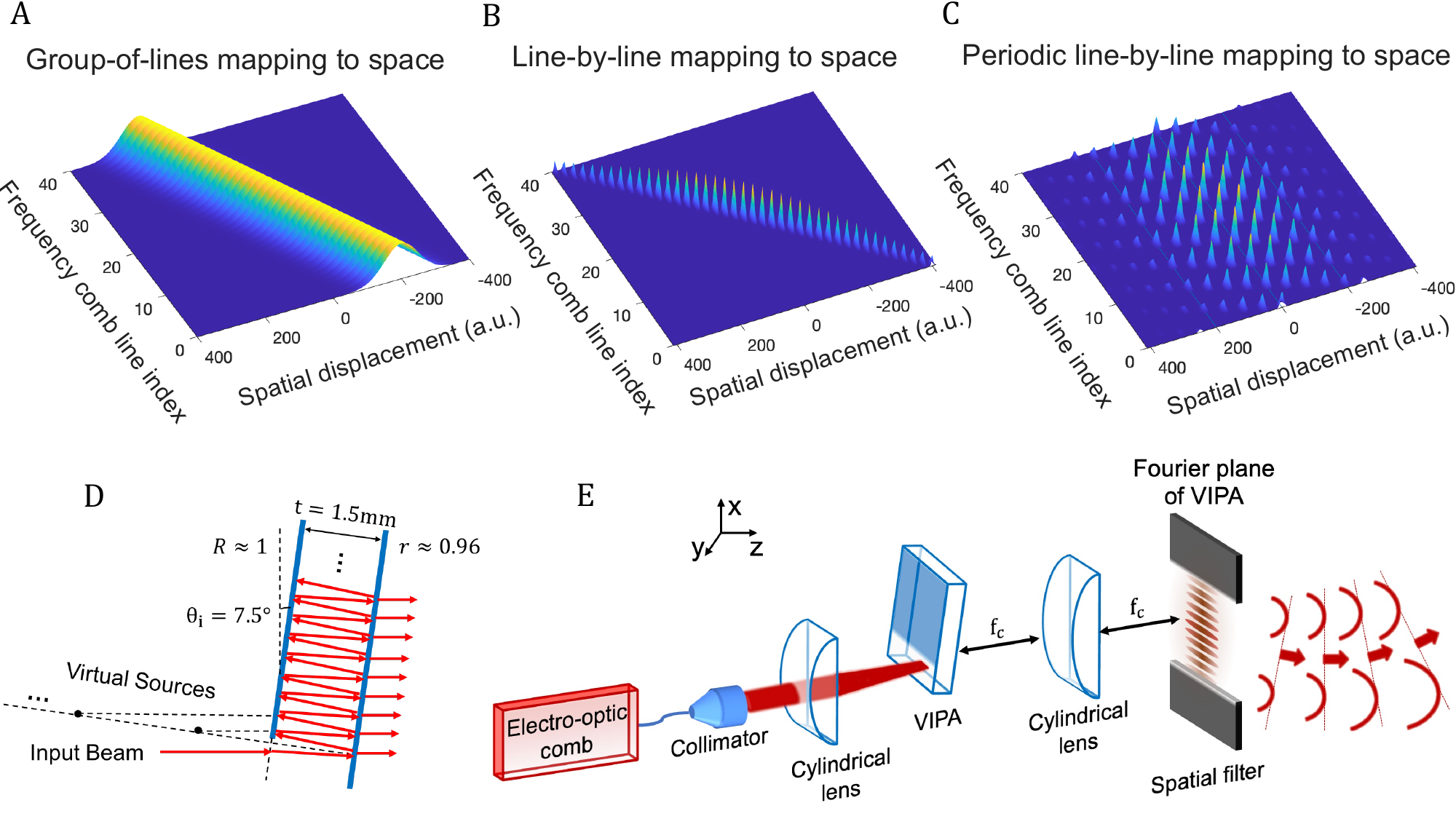}
\caption{ Simulated illustration of an optical frequency comb mapped to space such that---(A) the spatial spread of each comb line is wider than the spatial separation of consecutive comb lines, leading to multiple comb
lines grouped at each spatial location. The figure features 12 comb lines within the spatial full width at half maximum (FWHM) of each comb line; (B) The frequency comb lines are spatially resolved, that is, the spatial separation of consecutive comb lines is greater than the spatial spread of each comb line; (C) there is a periodic mapping of frequency comb lines to space. The comb lines are spatially discernible, however, each comb line periodically occupies multiple spatial locations. (D) VIPA structure, virtual sources, and multiple reflections within the VIPA. (E) Illustration of the experiment to realize a frequency-gradient array of individually-resolved comb lines. }

\label{fig1}
\end{figure*}

In the first segment of our work, we focus on realizing 100\% duty cycle steering thereby enhancing the time efficiency of dynamic beam control. Our methodology centers on achieving a fully resolved frequency-gradient array {[Fig~\ref{fig1}(B)]}, accomplished by utilizing the high repetition period of electro-optic (EO) frequency combs and the exceptional spectral resolution of a virtually imaged phased array (VIPA).

Building upon this foundation, we broaden the scope of our dynamic steering demonstration, transitioning from continuous-angular scanning to pulsed discrete-angular operation by leveraging the periodic spectral dispersion capabilities of the VIPA. We realize a frequency-gradient array composed of coherent optical frequency combs {[Fig~\ref{fig1}(C)]}, resulting in the dynamic scanning of broadband pulses. Additionally, we envision the on-chip demonstration of the presented schemes, 
and propose architectures incorporating full-duty-cycle steering while maximizing the benefits of integrated photonic implementation.%

\vspace{-6mm}
\section*{Full-duty-cycle steering}\label{100dutycycle}
\vspace{-3mm}

\begin{figure*}
  \centering
  \includegraphics[width=0.88\textwidth]{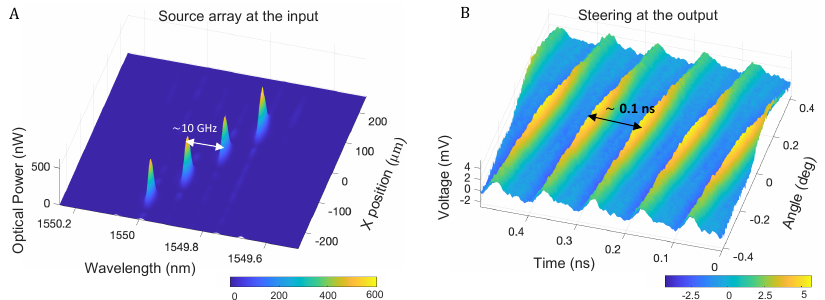}
\caption{{(A) Spectra measured in the Fourier plane of the VIPA indicating a frequency gradient array of four comb lines each spatially resolved. The comb lines are separated by 9.8 GHz in frequency and by $\sim$153 \textmu m in space. (B) Measured temporal waveforms from Spatial Fourier transform of (A) reveal full-duty-cycle continuous beam steering over an angular range of 0.8$^\circ$. The steering repeats every $\sim$101 ps consistent with the frequency spacing of the comb lines.}}
\label{fig2}
\end{figure*} 
In an OPA, a linear phase gradient is actively adjusted across the emitters to steer the beam. A frequency gradient array introduces a linear, time-dependent phase difference between emitters, resulting in a flat combined wavefront with continuously changing angular orientation over time, realizing dynamic beam steering. If the phase difference between consecutive array elements in an OPA is denoted as $\Delta \phi$, the corresponding time-dependent phase difference in a frequency gradient array with a frequency increment between consecutive elements of $\Delta \omega$ is expressed as $\Delta \phi = \Delta \omega t$. The emitted beam direction $\theta$ is given by~\cite{shaltout2019spatiotemporal}:

\begin{equation} \label{eq1}
     \sin\theta = \frac{\Delta \omega t + 2\pi m}{k_0 d}, m\in \mathbb{Z}
 \end{equation}

Here, $d$ represents the spacing between consecutive emitters, $k_0 = \omega_0/c$, where $\omega_0$ is the center frequency of the frequency gradient array, and $c$ is the speed of light in the medium. It is important to note that the steering angle varies with time without external control, and the angular steering speed is proportional to the frequency increment $\Delta \omega$.

Previous demonstrations~\cite{shaltout2019spatiotemporal, li2023picosecond} of frequency gradient arrays have illuminated mode-locked laser pulses onto a cascaded system of a diffraction grating and a lens. The grating maps frequencies to angles, and the lens performs a spatial Fourier transform, mapping the angles to space. 
These steering demonstrations have achieved a limited scan duration of 8 ps in a scan period of 12.5 ns~\cite{shaltout2019spatiotemporal}, and 81 ps in a scan period of 20 ns~\cite{li2023picosecond}, falling under a duty cycle of 0.5\%. 
We identify this limitation in the duty cycle of the scan as stemming from a poorly resolved frequency-gradient array composed of a group of comb lines mapped to each point in space, as illustrated in Fig.~\ref{fig1}(a). {The repetition rate of the mode-locked laser (below 100 MHz) used in the experiments was considerably lower than the spectral resolution of the diffraction grating (exceeding 10 GHz), 
which led to the spectral resolution spanning over a group of comb lines. For example, the duty cycle in the prior demonstrations can be explained by nearly 1500 comb lines~\cite{shaltout2019spatiotemporal} and 240 comb lines~\cite{li2023picosecond} spatially overlapping after spectral dispersion.} This results in steered waveforms that are isolated in time, and restricts the active scan time relative to the scan period given by $\Delta\omega^{-1}$. In our current work, we address this limitation by 
individually discerning frequency lines in space, as illustrated in Fig.~\ref{fig1}(B). Our approach is inspired by previous studies on high-resolution line-by-line Fourier pulse shaping, that achieved shaped-arbitrary waveforms with a full duty cycle via line-by-line spectral-phase and amplitude control of an input frequency comb. The shaped waveforms spanned the entire time-domain repetition period of the frequency comb resulting in high time-bandwidth products~\cite{ huang2008spectral, jiang2005spectral,cundiff2010optical}. In contrast, traditional group-of-lines pulse shaping generates low-duty cycle waveform~\cite{weiner2000femtosecond}.

Similarly, line-by-line resolved frequency-to-space mapping of the input frequency comb can result in steering spanning the entire scan period achieving 100\% duty cycle. To illustrate this concept, in our experiments, we use an electro-optic (EO)
comb as a source, with a repetition rate of approximately 9.8 GHz. This comb is generated through a series of intensity and phase modulations applied to a continuous wave laser, resulting in {spectrally-flat, evenly-spaced} sidebands around the carrier frequency~\cite{metcalf2013high}. Subsequently, the EO comb is directed to a virtually imaged phased array (VIPA), a Fabry-Perot etalon with a side entrance renowned for its high-resolution spectral dispersion capabilities~\cite{xiao2004dispersion, xiao2005experimental}. As illustrated in Figs.~\ref{fig1}(D), the VIPA consists of two plane parallel mirrors separated by a glass material. The front mirror exhibits nearly 100\% reflectivity, except for a specific window area used to couple light into the device, and the back mirror is partially reflective, typically with over 95\% reflectivity. The input beam is focused into the VIPA by a cylindrical lens at an incident angle of $7.6^\circ$. The beam bounces back and forth between the two mirrors, producing multiple reflections on the back mirror of the VIPA. The transmitted beams from the back mirror appear as though they originate from an array of virtual sources spaced from their corresponding reflection spots by the distance traveled by the beam within the VIPA~\cite{shirasaki1996large}. The linear delay gradient across the transmitted beams generates a frequency-dependent phase difference and induces angular dispersion as the beams diffract and interfere. The VIPA is followed by a cylindrical lens that maps the frequencies from angle to space along the dispersion axis~(x-axis) on the back focal plane or the Fourier plane of the VIPA. Additional information regarding the experimental setup and the VIPA's specifications can be found in~\cite{MaterialsMethods}.

{The VIPA} surpasses conventional diffraction gratings in angular dispersion, exhibiting significantly enhanced resolving power as the number of virtual sources in the array increases. This characteristic enables VIPAs to achieve sub-GHz resolutions, outperforming regular diffraction gratings by one to two orders of magnitude.

Similar to Fabry-Perot etalons, the VIPA possesses a free-spectral range (FSR) determined by the optical path length between two reflections. In the case of a broadband input, frequencies separated by the VIPA-FSR are directed to the same angle and spatial position on the Fourier plane, as illustrated in Fig.~\ref{fig1}(c). This presents a challenge in applications like spectral filtering and pulse shaping, where the total bandwidth can be constrained by the VIPA-FSR (typically within a few hundred GHz), and hence another broadband disperser is often introduced along the orthogonal axis to resolve this ambiguity~\cite{supradeepa2008femtosecond,lee2023sub}. On the other hand, in this work we exploit the periodic dispersion feature of the VIPA to our advantage in subsequent experiments, unlocking the capability for discrete-angular dynamic steering.

\begin{figure*}
  \centering
  \includegraphics[width=0.88\textwidth]{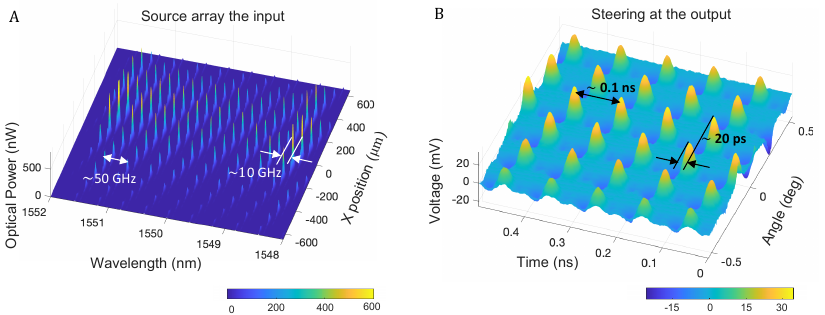}
 \vspace{-0.2in}
\caption{{(A) Spectra measured in the Fourier plane of the VIPA unveil a spatially and spectrally resolved frequency-gradient array of broadband frequency combs. This array features a periodic frequency-to-space mapping occurring every 49.2 GHz as dictated by the FSR of the VIPA. The consecutive comb lines that are spatially displaced exhibit a separation of 9.8 GHz in frequency and approximately 153 \textmu m in space. (B) Measured temporal waveforms from Spatial Fourier transform of (A) showcase full-duty-cycle pulsed beam steering over an angular range of 1$^\circ$. The steering period is $\sim$101 ps, consistent with the repetition rate of the input comb. The consecutive angular peaks of the pulses exhibit a delay of $\sim$20.4 ps consistent with the FSR of the VIPA.}}
\label{fig3}
\end{figure*}

In our experiments, we employ a VIPA with an FSR of $\sim$49.2 GHz at an input angle of 7.6$^\circ$. The line-spacing of the EO comb is set to $\sim$9.8 GHz.
A pulse shaper is used to limit the total number of comb lines that are routed into the VIPA to four. {A spatial filter in the Fourier plane selects the lowest diffraction order of the VIPA.} The frequency spectrum on the Fourier plane is plotted in Fig.~\ref{fig2}(A). The selected single diffraction order contains four comb lines spatially separated by $\sim$91.8 µm, consequently, the frequency-array is characterized by a linear gradient of $\Delta \omega/d = 2\pi \times 106.9$ MHz/um. 
We observe that the individual comb lines in the array are discernible in space. 
The theoretical resolution of the VIPA is expected to be under 1 GHz~\cite{xiao2005experimental}, significantly surpassing the 9.8 GHz-repetition rate of the EO comb.

To emulate the far-field pattern produced by the frequency-gradient array, we employ spatial Fourier transform using a spherical lens following the Fourier plane~\cite{MaterialsMethods}. The steered beam at the output is coupled into a single-mode fiber, with subsequent detection utilizing a 23 GHz-bandwidth photodetector. The time-domain waveforms are then observed on a sampling scope and plotted as a function of angle in Fig.\ref{fig2}(B). The waveforms reveal continuous dynamic steering of the beam over a range of $0.8^\circ$ with a scan rate of 9.8 GHz, as dictated by Eq.~(\ref{eq1}). This demonstration boasts a duty cycle of 100\%, a direct outcome of resolving the individual comb lines within the frequency gradient array.

\vspace{-6mm}
\section*{Pulsed beam steering}\label{PulsedBeamSteering}
\vspace{-3mm}

\begin{figure*}
  \centering
  \includegraphics[width=0.92\textwidth]{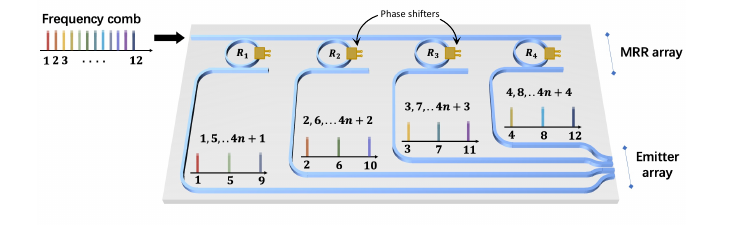}
 \vspace{-0.2in}
\caption{{Illustration of an on-chip implementation design to realize a frequency gradient array for full-duty cycle beam steering. The design features a microring filter array used to select desired set of comb lines from the input comb that are subsequently routed to emitters.}}
\label{fig4}
\end{figure*}

Advancing beyond the continuous steering demonstration, we explore broadband emitters positioned in the Fourier plane of the VIPA designed for pulsed discrete-angular dynamic steering. An EO comb with an FSR of $\sim$9.8 GHz is generated featuring a flat spectrum with $\sim$ 50 comb lines within a 10 dB bandwidth. This comb is seeded into the VIPA, which has an FSR of $\sim$49.2 GHz. The comb line-spacing is chosen to precisely match an integer fraction of the VIPA FSR, resulting in subsets of comb lines in the Fourier plane of the VIPA as measured and plotted in Fig.\ref{fig3}(A). The comb lines periodically superimpose on each other every 49.2~GHz (VIPA-FSR), and the adjacent subsets of comb lines feature an offset frequency shift of 9.8~GHz (comb repetition rate). In the absence of a spatial filter, the spatial distribution of the input comb lines measured in the VIPA's Fourier plane is plotted in Fig.\ref{fig3}(A). The frequency-to-space mapping is fairly linear explained by the dispersion law of VIPA~\cite{MaterialsMethods}. While periodic dispersion of the VIPA is often perceived as a compromise on the total bandwidth, in our approach, we exploit this characteristic to dynamically scan broadband pulses.
The time-domain waveforms measured after the spatial Fourier transform of the VIPA's Fourier plane is presented in Fig.\ref{fig3}(B). The steered pulse train peaks at discrete angles spaced by 0.188$^\circ$. The pulses exhibit a period of $\sim$101$\pm$0.2 ps at a fixed angular orientation consistent with the comb repetition rate. In other words, the angular scan rate is determined by the carrier-envelope offset increment of 9.8 GHz between adjacent combs in the Fourier plane of the VIPA. The pulses are delayed by $\sim$20.4 $\pm$ 0.2 ps between consecutive angular peaks, in line with the FSR of the VIPA. The active scanning time spans the entire scan period, affirming a full-duty-cycle operation. The theoretical prediction in close agreement with the experiment is detailed in the supplementary text S1. Note that the angular scanning observed here can also be understood on the basis of the fundamental relationship between spectral dispersion and transverse delay-gradient across the spatial aperture~\cite{weiner2000femtosecond,weiner2011ultrafast}.


\vspace{-4mm}
\section*{Potential for on-chip implementation}\label{PIC}
\vspace{-3mm}

We outline a design for an integrated photonic implementation for achieving both continuous and discrete angular dynamic beam steering with a full-duty cycle.
The frequency-gradient array can be realized using microring filter arrays by leveraging their resolution at a single GHz-level or lower. An illustration of the design is shown in Fig.~\ref{fig4} where microring resonators (MRRs) with nearly identical free-spectral range are serially coupled to a bus waveguide. Similar designs with higher-order MRR filters have been explored to realize on-chip wavelength selective switches using silicon photonics technology ~\cite{cohen2023fine}. An EO frequency comb is seeded to the bus waveguide with its repetition rate chosen to be an integer fraction of the microring's FSR. Fig.~\ref{fig4} illustrates an example where MRR-FSR is four times the line-spacing of the input frequency comb. Through tuning of MRR resonances using thermo-optic phase-shifters, the resonance spacing between consecutive MRRs is set to the input comb's line-spacing, facilitating the selection of a subset of input comb lines at the drop ports of the MRRs, referred to as sub-combs hereafter. These sub-combs exhibit a repetition rate matching the MRR-FSR, and they are linearly frequency-shifted by the input comb's line-spacing. The resulting frequency-gradient array of sub-combs is then directed to an array of edge emitters, enabling dynamic steering in the far field.

Similar to OPA systems, the total number of emitters and their spacing dictates the beam width and the field-of-view.  The quality factor and resonance linewidth of the MRR can be chosen to accommodate differences in the FSRs of the MRRs due to fabrication imperfections.  The cumulative path lengths traversed by each subcomb through the bus waveguide, MRR, and drop ports are matched (to well within the pulse duration) enabling time-synchronized emission of the pulses. Notably, this design can be applied to both continuous and pulsed dynamic scan scenarios by appropriately adjusting the input pump bandwidth. 

An alternative on-chip realization involves using a spectral disperser, such as the Arrayed Waveguide Grating (AWG)~\cite{yoshikuni2002semiconductor, cheben2007high}. The AWG produces periodic dispersion at its output, which can be employed for beam steering, akin to our experiments with a VIPA. Recent advancements in AWG design have showcased enhanced spectral resolution at the GHz-scale~\cite{gatkine2017arrayed, zhang2022arrayed, gatkine2023near, stoll2020performance}. Both the MRR-based and AWG-based implementation approaches eliminate the need for real-time phase tuning to control the output beam orientation. However, the MRR-based approach stands out by featuring notably narrow linewidths and reconfigurable resonances, providing the capability to adapt to different scan rates using a single device.

\vspace{-3mm}
\section*{Summary and conclusions}\label{Summary}
\vspace{-3mm}

This work illustrates how to realize time-efficient dynamic scanning that achieves a 100\%-duty cycle by resolving individual lines in the frequency-gradient source array. By further leveraging the operational characteristics of the VIPA, we have presented a design for a broadband source array capable of achieving dynamic pulsed beam steering.    
Our demonstration achieves scanning in discrete steps of {0.188$^\circ$} and {$\sim$20.4 ps} in the angular and temporal domains, respectively. 

We explore on-chip implementations for dynamic steering schemes that can harness the merits of integrated photonic technology, including miniaturization, wide field-of-view operation, and high spatial resolution. Notably, the presented techniques eliminate the need for real-time tuning of phase, delay, or wavelength to steer the beam. This work opens avenues for versatile dynamic beam control with the potential for comprehensive angle, range, and time-resolved imaging.


\newpage

\bibliography{References}
\vspace{-6mm}
\section*{Acknowledgements}\label{Acknowledgements}
\vspace{-3mm}
We thank Dr.~Jason~D.~McKinney for valuable discussions and Avanex Corporation for the VIPA devices. \textbf{Funding:} This work was supported by the Air Force Office of Scientific Research grant ({FA9550-20-1-0283}). \textbf{Author contributions:} S.S. developed the concepts, conducted the theoretical analysis, and assisted in the experiments. J.W. conducted the experiments. A.M.W. developed the concepts and supervised the project. All the authors contributed to and approved the manuscript.  \textbf{Competing interests:} The authors declare no other competing interests. \textbf{Data and materials availability:} All data are available in the main text or the supplementary materials. 
\vspace{-5mm}
\section*{Supplementary materials}\label{Supplementary materials}
\vspace{-3mm}
\noindent {Materials and Methods \\
Figs. S1 to S8 \\
Table S1\\
References \textit{\cite{ lancis2008lossless, supradeepa2010self}}}
\end{document}


\title{\large{Supplementary material: Ultrafast dynamic beam steering with optical frequency comb arrays}}

\maketitle
%




\section*{Materials and Methods}\label{methods}

\subsection*{VIPA characterization}\label{VIPA}

\begin{figure*}
  \centering
  \includegraphics[width=\textwidth]{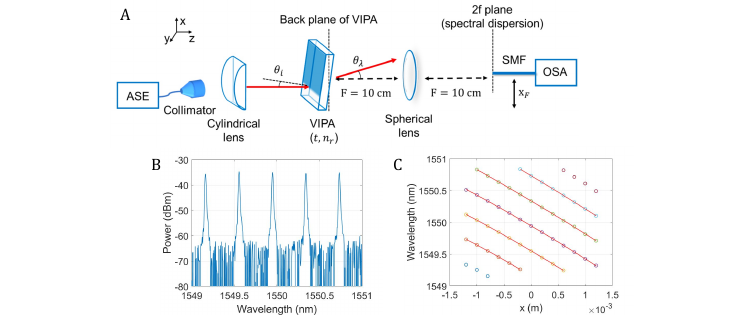}
\caption{(A) Experimental setup for VIPA characterization. ASE: Amplified spontaneous emission, VIPA: Virtually imaged phased array, SMF: Single-mode fiber, OSA: Optical spectrum analyser. (B)~Spectrum measured at one spatial location in the Fourier plane of VIPA. (C) Dispersion data from measurements in the Fourier plane of the VIPA resulting in a fitted incident angle of 7.56$^\circ$.}
\label{fig_2fsetup}
\end{figure*}

A solid VIPA with a nominal thickness ($t$) of 1.5 mm
and a nominal refractive index ($n_r$) of 2, is used in our experiments, which corresponds to a 50 GHz FSR. The experimental setup for characterizing the VIPA is shown in Fig.~\ref{fig_2fsetup}(A). The angular dispersion behavior of the VIPA is characterized by using a broadband source as input and measuring the spectra at different spatial positions on its Fourier plane. The incident angle of the beam entering VIPA is determined using the dispersion law of the VIPA~\cite{xiao2004dispersion}. In this setup, the input source is an amplified stimulated emission (ASE) source with a bandwidth ranging from 1530 nm to 1570 nm. The collimated beam from the input source is focused into the VIPA by a cylindrical lens. For precise control of the incident angle, the VIPA is mounted on a fine rotational stage. A spherical lens with a focal length of F = 10 cm focuses light from different output angles onto different positions on its back focal plane. The spectrum is measured while scanning a single-mode fiber (SMF) along the dispersion axis (x-axis) on the back focal plane of the lens. An example of the output spectrum is presented in Fig.~\ref{fig_2fsetup}(B). The translation stage used for scanning has a resolution of 0.01 mm, and the mode diameter of the SMF is also $\sim$0.01 mm.  The spectrum shows multiple peak intensities separated by approximately 0.4 nm, in accordance with the nominal 50 GHz FSR. The FWHM of the peaks is measured to be approximately $\sim$0.02 nm (2.5 GHz), equal to the resolution of the OSA used in the experiment, while the theoretical minimum full-width at half maximum (FWHM)~\cite{xiao2005experimental} is calculated to be $\sim$5 pm (600 MHz). The position of the fiber is carefully adjusted to be on the Fourier plane to prevent any broadening and asymmetry caused by the spatial chirp effect. Then we found the peak output wavelengths at different spatial positions from the spectra and fit them using the dispersion law. The dispersion law for a solid VIPA can be written as~\cite{xiao2004dispersion,xiao2005experimental}

\begin{equation}\label{Supp_eq1}
\lambda_p = - \frac{\lambda_0}{2n^2_rF^2}x^2_F - \frac{\lambda_0\tan\theta_{in} \cos\theta_i}{F n_r \cos\theta_{in}}x_F + \lambda_0
\end{equation}

\noindent Here $x_F$ is the output position along the dispersion axis of the Fourier plane, $\theta_i$ is the incident angle, $\theta_{in}$
is the angle inside the VIPA after the beam gets refracted while traversing from air to glass, where, $\sin(\theta_i) = n_r \sin(\theta_{in})$. The zero $x_F$ position is identified by finding the position with maximum output power. The experimental data from the Fourier plane of the VIPA is fit to the above second-order polynomial, and in the process the incident angle and the focal length of the spherical lens are estimated. Figure~\ref{fig_2fsetup}(C) presents the dispersion data measured. The experimental data points are represented by circles, while the fitting results are shown as lines. The spectra were measured with a step length of 0.2 mm within a range of 2 $\sim$ 3 mm. The focal length estimated from the fit is 10.11$\pm$ 0.74 cm, 
in close agreement with the nominal value of 10 cm. The incident angle is estimated to be 7.56 $\pm$ 0.56 degrees. Here, the angle is large enough to allow the linear term in the dispersion equation to dominate over the quadratic resulting in near-uniform frequency-to-space mapping.

\begin{figure*}
  \centering
  \includegraphics[width=\textwidth]{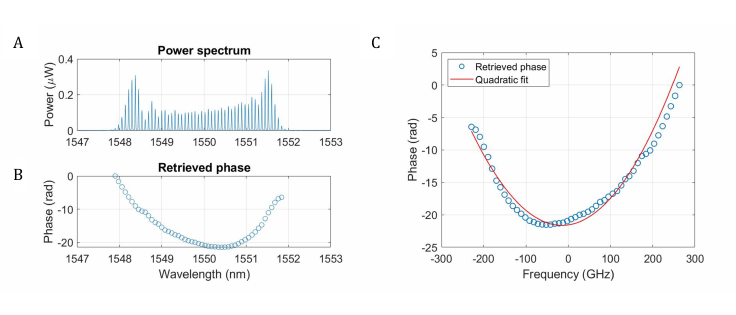}
\caption{(A) Spectrum of the input comb used in pulsed-steering experiments (when three PMs are cascaded in the EO-comb generator). (B) Retrieved spectral phase of the input comb. (C) Quadratic fit for the retrieved spectral phase. }
\label{fig_6_EOSpectrumPhase}
\end{figure*}

\subsection*{Electro-optic comb characterization}\label{EO-comb}

The electro-optic comb in our experiment is generated as described in \cite{metcalf2013high}. The setup consists of three phase modulators (PMs) and one intensity modulator (IM), all driven by the same RF oscillator. The IM ($V_{\pi}\sim$ 5.5 V at 10 GHz) is used to generate a pseudo-square pulse train by appropriately adjusting its bias.  In order to achieve a broad spectrum in pulsed steering experiments three PMs are cascaded. Each PM has $V_{\pi}\sim$ 3.5 V at 10 GHz and can handle a maximum RF power of 1 W, hence adding more than 20 lines within -10 dB bandwidth. When the RF sinusoids driving the modulators are correctly aligned in time such that the flat-top pulse train only carves the modulated light at the peaks or valleys of the sinusoidal phase modulation, a relatively flat spectrum can be obtained. The applied phase can be approximated to be quadratic at the peaks/valleys of sinusoidal phase modulation. This results in time-to-frequency mapping due to time-lensing effect where the spectral envelope takes the shape of the pseudo-square temporal envelope of the input pulse. The generated EO comb spectrum when three PMs are included with the IM is shown in Fig.~\ref{fig_6_EOSpectrumPhase}(A). At the outer edges of the spectrum in this EO frequency comb setup, characteristic "rabbit ears" can be observed which stem from the distortion of the parabolic phase modulation (or time lens aberrations~\cite{ metcalf2013high, lancis2008lossless}) within the intensity-modulated temporal pulse profile.


\begin{figure*}[b]
  \centering
  \includegraphics[width=\textwidth]{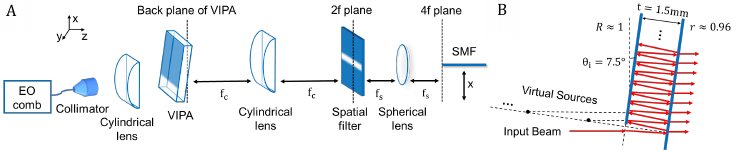}
\caption{(A) Experimental setup. MLL: mode-locked laser, VIPA: virtually imaged phased array, SMF: single-mode fiber. (B) VIPA structure, virtual sources, and multiple reflections within the VIPA.}
\label{fig_4fsetup}
\end{figure*}

The generated EO comb has a residual near-parabolic spectral phase which needs to be compensated in order to compress the temporal pulses to their bandwidth-limited duration.
By compensating for this quadratic phase using a pulse shaper, transform-limited picosecond pulses can be obtained~\cite{metcalf2013high}. 
We adopted a 4f lens system to achieve steering in our experiments as shown in Fig.~\ref{fig_4fsetup}(A) and detailed in the next subsection. 
To retrieve the spectral phase of the pulses steering at the output of our experimental setup,
we adopted a self-referenced and easy-to-implement technique based on dual-quadrature spectral interferometry~\cite{supradeepa2010self}. 
Subsequently, we compensated for the retrieved phase using a pulse shaper, effectively compressing the steered pulses. The retrieved spectral phase of the EO comb is plotted in Fig.~\ref{fig_6_EOSpectrumPhase}(B). As expected, there is a dominant quadratic phase term present. The quadratic fit for the spectral phase is plotted in Fig.~\ref{fig_6_EOSpectrumPhase}(C), with a quadratic coefficient of $3.17 \times 10^ {-1}$ radians/GHz$^2$.

\begin{figure*}
  \centering
  \includegraphics[width=\textwidth]{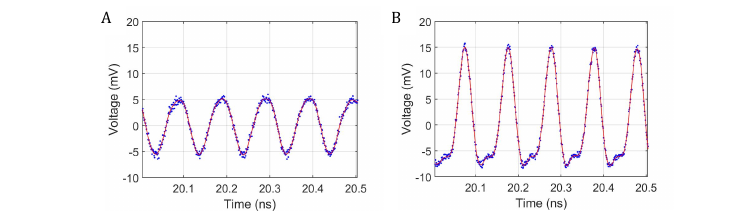}
\caption{Temporal profile of the EO comb pulses measured at the output plane of the experiment: before~(A) and after~(B) pulse compression. }
\label{fig_7_EOTemporalPulse}
\end{figure*}

To compress the steered pulses, we used a commercial pulse shaper to compensate for the retrieved quadratic phase. The linear term of the retrieved phase was neglected since it corresponds to a constant delay in the time domain. Similarly, the constant phase was also disregarded. The steered pulses on the 4f plane were detected using a 23 GHz photodetector and measured on a sampling scope. The measured pulse waveform before and after compression is shown in Fig~\ref{fig_7_EOTemporalPulse}(A,B). After compression, the peak intensity doubled, and the measured pulse duration narrowed to $\sim$30 ps. The measured pulse duration is limited by the bandwidth of the detector, and the actual pulse duration can be accurately determined using autocorrelation techniques. For the case of a transform-limited pulse where the spectral phase is perfectly compensated, the calculated pulse duration is approximately 2 ps, corresponding to the $\sim$500 GHz bandwidth of the comb.

\begin{figure*}
  \centering
  \includegraphics[width=\textwidth]{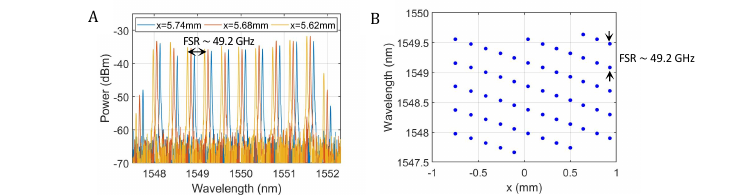}
\caption{(A) Spectra of three adjacent frequency combs in the Fourier plane of the VIPA. (B) Spatial distribution of the comb teeth in the Fourier plane of the VIPA.}
\label{fig_8_EO2fplane}
\end{figure*}

When the EO comb is seeded into the VIPA and the Fourier plane is measured using the setup in Fig.~\ref{fig_2fsetup}, we fine-tuned the comb line-spacing around 9.7-10 GHz to set it as an integer fraction of the VIPA FSR such that each subset of comb lines overlapped at the same position. The comb line-spacing is set to $\sim$9.8 GHz in the above manner, and the measured spectra of three adjacent frequency combs on the 2f plane are plotted in Figure~\ref{fig_8_EO2fplane}(A). Figure~\ref{fig_8_EO2fplane}(B) shows the spatial distribution of each comb line from the input EO comb. Each dot represents a frequency mode, and each row of dots is tilted according to the angular dispersion of VIPA. The dots are separated by a spatial spacing of 153 $\pm$ 9 \textmu m, and the tilt corresponds to a frequency shift of 9.81 $\pm$ 0.07 GHz. The VIPA FSR given by the separation between the columns is 49.16 $\pm$ 0.03 GHz.




\subsection*{Experimental setup}

The input beam is focused into the VIPA using a cylindrical lens with a focal length of 10 cm. The incident angle is found to be $\sim$$7.56^\circ$ using dispersion law of the VIPA as described earlier by Eq.~(\ref{Supp_eq1}). In the steering experiment, the output beam from the VIPA then undergoes frequency-to-space mapping by a cylindrical lens with a focal length $f_c$ of $\sim$6 cm, resulting in a spatial array of frequency-shifted spectra along the dispersion axis (x-axis) on the Fourier plane or the 2f plane. The setup is shown in Fig.~\ref{fig_4fsetup}. To examine the far-field pattern, a spherical lens with a focal length $f_s$ of around 3.5 cm is introduced after the 2f plane. The lens maps the steering angle to a position on the x-axis at its back focal plane or the 4f plane. Since the two lenses form a 4f imaging system, the 4f plane will represent a scaled-down version of the VIPAs back plane, with a demagnification factor of $\sim$0.58 along the x-axis. Along the y-axis, the collimated output beam from the VIPA is focused by the spherical lens.

\begin{figure*}
  \centering
  \includegraphics[width=\textwidth]{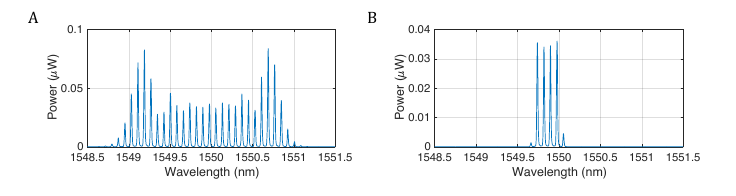}
\caption{(A) Spectrum of the input comb used in continuous steering experiments (with a single PM engaged in the EO-comb generator) (B) Four comb lines selected using a pulse shaper in the continuous steering experiments and seeded into the VIPA.}
\label{fig_8_FourEOComblines}
\end{figure*}

\begin{figure*}
  \centering
  \includegraphics[width=\textwidth]{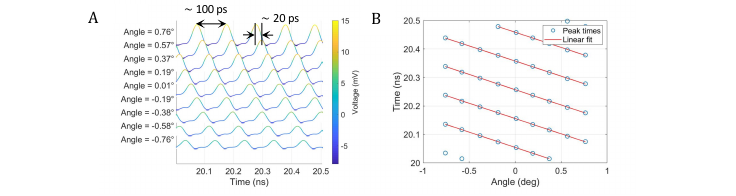}
\caption{(A) Temporal waveforms of the steered pulse trains measured on the 4f plane and plotted as a function of steering angle. (B) Pulse peak times of the steered pulse train}
\label{fig_delayGradient}
\end{figure*}

The results measured at the Fourier plane (or 2f plane) of the VIPA shown in the main text in figures~2(A) and~3(A) are initially obtained using the setup depicted in Fig.~\ref{fig_2fsetup} with a spherical lens of focal length F=10 cm. They are subsequently scaled (by a factor of 6/10) before plotting to represent the dimensions in the subsequent steering experiment, which employed a cylindrical lens of focal length $f_c=6$ cm before the Fourier plane of the VIPA. All measurements presented in Fig.~2 and Fig.~3 of the main text were acquired using an SMF~(with mode-diameter 10 \textmu m), translated in the 2f and 4f planes utilizing a computer-controlled Piezo Actuator (Newport's NanoPZ actuator PZA12) with a resolution under 1~\textmu m. 


The back plane of the VIPA comprises an array of delayed beams. The delay corresponds to the separation of the virtual sources and can be measured using the VIPA FSR found in the 2f plane experiments. For our setup, we expect the delay to be around 20 ps due to the $\sim$50 GHz VIPA FSR. To observe the delayed beams, a single-mode fiber (SMF) connected to a 23 GHz photodetector (PD) is scanned along the x-axis in the 4f plane.

In the continuous steering experiment (Fig.~2 in the main text), only one PM is engaged to obtain about 20-30 comb lines in 10-dB bandwidth as plotted in Fig.~\ref{fig_8_FourEOComblines}(A). A pulse shaper is used to select four comb lines from this spectrum as shown in Fig.~\ref{fig_8_FourEOComblines}(B) which is then routed into the VIPA. Subsequently, in the Fourier plane of the VIPA, only a single diffraction order is selected by employing a spectral filter. The experimental conditions used in the continuous and discrete-angular scanning experiments are organized in Table~\ref{table1}.

The focal lengths of the cylindrical and spherical lens chosen in the 4f lens system result in a beam width of $\sim$30 \textmu m along x-axis and $\sim$12 \textmu m along y-axis in the 4f plane.
The pulses in the 4f plane were probed using an SMF~(with mode-field diameter $\sim$10 \textmu m) and detected using a 23 GHz PD and measured on a sampling scope. A 10 MHz reference RF signal provided by the RF oscillator used for EO comb generation was used as the trigger for the sampling scope. The pulse waveforms were averaged 64 times on the scope and then curve-fitted. All of the pulse trains were well compressed with a measured FWHM duration of $\sim$30 ps. This indicated that a similar amount of residual chirp was present in all of the steered pulses after compensating for the quadratic phase.


In discrete-angular steering experiments, the steered pulse trains exhibit a uniform spatial distribution with a spacing of $\sim$115 µm (translating to an angular separation of 0.19$^\circ$). This is in agreement with the demagnification factor ($\sim$0.58) of the 4f system, where the calculated beam separation on the back plane of the VIPA at an incident angle of 7.56$^\circ$ is $\sim$196 \textmu m. As shown in Fig.~\ref{fig_delayGradient}(A), each pulse train has a time period of $\sim$100 ps and is delayed by $\sim$20 ps with respect to each other. Figure~\ref{fig_delayGradient}(B) illustrates the pulse-peak times of all pulse trains, indicating a linear gradient in their delays. The fitted delay, which corresponds to the temporal pulsing rate, was 20.4$\pm$0.2~ps, consistent with the FSR of the spatial array of frequency combs on the 2f plane. The time period, indicating the beam scanning rate, was averaged to be 101$\pm$0.2~ps, aligning with the frequency gradient of the frequency comb array.

\begin{center}

\begin{tabular}{ | m{5em} | m{4.8cm}| m{4.3cm} | m{4cm} |} 
  \hline
 \hspace{9em} & EO comb generation & Spectral filtering of the EO comb  & Spatial filtering in VIPA's Fourier plane \\ 
  \hline
Continuous-angular steering & Single PM cascaded with IM to generate the comb with about 25 comb lines in 10-dB bandwidth. & A pulse shaper is used to select 4 comb lines from the EO comb that are seeded into the VIPA. & Spatial filter employed to select single diffraction order of the VIPA. \\ 
  \hline
Discrete-angular steering & Three PMs cascaded with IM to generate the comb with about 50 comb lines in 10-dB bandwidth.  & No spectral filtering employed. & No spectral filtering employed. \\ 
  \hline
\end{tabular}
\captionof{table}{Experimental conditions in the continuous- and discrete-angular steering experiments.}\label{table1}
\end{center}

\begin{figure*}
  \centering
  \includegraphics[width=\textwidth]{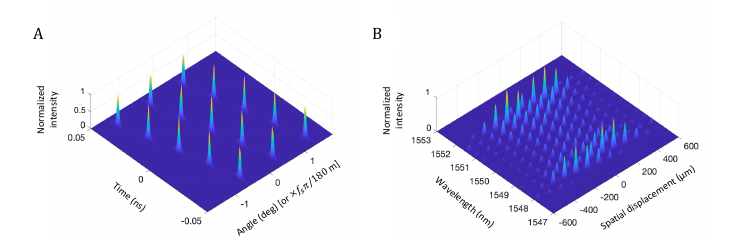}
 \vspace{-0.2in}
\caption{Theoretical simulation: (A) spatial array of delayed combs, and (B) spatial Fourier transform of (A).}
\label{fig_simulation}
\end{figure*}

\subsection*{Numerical simulation of the frequency-gradient array for pulsed beam steering}\label{S1}



In this section we relate the pulsed-beam steering at the output, (that is, 4f plane in Fig.~\ref{fig_4fsetup} of the experiment) to the periodic array of frequency gradient combs in the 2f plane. We represent the beam steering at the output with a spatial array of frequency combs, each with a repetition rate of 9.81 GHz and delayed with respect to each other by 20.4 ps. Here the frequency combs are approximated to be Gaussian beams with uniform intensity and uniform widths($\sim$ 30 \textmu m). Figure~\ref{fig_simulation}(A) shows the time-domain representation considered over a single scan period, consisting of 15 frequency combs separated by 115 \textmu m or 0.188$^\circ$ in angle (considering the spherical lens of focal length $f_s =$ 3.5 cm in the setup Fig.~\ref{fig_4fsetup}). The comb spectrum in Fig.~\ref{fig_6_EOSpectrumPhase}(A) with constant spectral phase is used in the simulation. The spatial Fourier transform of such a comb array by the spherical lens (of focal length $f_s$) is computed and plotted in Fig.~\ref{fig_simulation}.  The comb lines are separated by 94.3 \textmu m and span a spatial range of approximately over 1200 \textmu m. This is in close agreement with the experimental measurements within the measurement precision.



\section*{References}
\vspace{0.1in}
\begingroup
\renewcommand{\section}[2]{}%
\vspace{-0.4in}